\documentclass[epj]{svjour}

\usepackage{graphicx}
\usepackage{fancyhdr}
\def\makeheadbox{{
 }}
\def\mytitle{My title} 
\def\myauthors{My name}  
\def\mytype{My type of session}
\def\mysession{My session}
\def\mytitle{Large $\tan\beta$ effects in flavour physics} 
\def\myauthors{Gino Isidori}    
\def\mytype{Review}
\def\mysession{\myauthors}
\newcommand{\be}{\begin{equation}}
\newcommand{\ee}{\end{equation}}
\newcommand{\bea}{\begin{eqnarray}}
\newcommand{\eea}{\end{eqnarray}}
\newcommand{\beq}{\begin{equation}}
\newcommand{\eeq}{\end{equation}}
\newcommand{\ba}{\begin{array}}
\newcommand{\ea}{\end{array}}
\newcommand{\beqa}{\begin{eqnarray}}
\newcommand{\eeqa}{\end{eqnarray}}

\newcommand{\cL}{{\cal L}}

\newcommand{\cA}{{\cal A}}
\newcommand{\cO}{{\cal O}}

\newcommand{\no}{\nonumber}

\newcommand{\lsim}{\stackrel{<}{_\sim}}
\newcommand{\gsim}{\stackrel{>}{_\sim}}

\newcommand{\BR}{{\mathcal B}}
\newcommand{\cB}{{\mathcal B}}

\newcommand{\Btaun}{{B \to \tau \nu}}

\def\plb#1#2#3{    { Phys. Lett. }{\bf B #1} (#2) #3}

\def\prd#1#2#3{    { Phys. Rev. }{\bf D #1} (#2) #3}

\def\jhep#1#2#3{   { JHEP  }{\bf #1} (#2) #3}

\begin{document}

\title{\boldmath Large $\tan\beta$ effects in flavour physics}

\author{Gino Isidori
\thanks{\emph{Email:} Insert  Email  of corresponding author here}%
}                     % Do not remove
\institute{INFN, Laboratori Nazionali di Frascati, Via E. Fermi 40 
           I-00044 Frascati, Italy}
\date{}
\abstract{
After a short introduction to the SUSY flavour problem, 
we focus the attention on the MSSM with MFV and large $\tan\beta$.
The theoretical motivations and the general features 
of this scenario are briefly reviewed. 
The possible signatures in 
low-energy flavour-violating observables are discussed, 
with particular attention to the role played by  
$\BR(P\to\ell\nu)$, $\BR(B_{s,d}\to \ell^+\ell^-)$
and $\BR(\mu\to e \gamma)$.}

%     } % end of PACS codes
% end of abstract
%
%
% Remember the following: Page limits: 10 p for plenary talks
%
\maketitle

\section{Introduction: the SUSY flavour problem and the MFV hypothesis}
In most extensions of the Standard Model (SM), including the 
so-called MSSM (Minimal Supersymmetric SM),
the new degrees of freedom which modify the ultraviolet 
behavior of the theory appears only around or above
the electroweak scale ($v \approx 174$ GeV). 
As long as we are interested 
in processes occurring below this scale (such as $B$, $D$ and $K$ 
decays), we can integrate out the new degrees of freedom 
and describe the new-physics effects --in full 
generality-- by means of an Effective Field Theory (EFT) approach.
The SM Lagrangian becomes the renormalizable part of a more general 
local Lagrangian which includes an infinite tower of higher-dimensional 
operators construc\-ted in terms of SM fields. The higher-dimensional 
operators are suppressed by inverse powers of a dimensional parameter, 
the effective scale of new physics, which in the MSSM case 
can be identified with the mass scale of the soft-breaking terms. 

This approach allows us to analyse all realistic 
extensions of the SM in terms of a limited number of parameters.
In particular, it allows us to investigate the  
flavour-symmetry breaking pattern of the model
without knowing the dynamical details of the theory 
above the electroweak scale. 
In case of a generic flavour structure, the higher-dimensional operators 
should naturally induce large effects in processes which are not mediated 
by tree-level SM amplitudes, such as $\Delta F=1$ and  $\Delta F=2$
flavour-changing neutral current (FCNC) transitions. 
Up to now there is no evidence of these effects and this 
implies severe bounds on the effective scale of new physics.
For instance the good agreement between SM 
expectations and experimental data on $K^0$--${\bar K}^0$, 
$B_d$--${\bar B}_d$, and $B_s$--${\bar B}_s$ mixing amplitudes 
leads to bounds above $10^4$~TeV, $10^3$~TeV, and $10^2$~TeV, 
respectively.
In the MSSM, where the new degrees of freedom are expected to be 
around the TeV scale, these bounds represent 
a serious problem:  if we insist that squarks in the TeV range 
are necessary for a stabilization of the Higgs sector, 
we have to conclude that the model has a highly non-generic 
flavour structure, similar to the SM one.

The quark-flavour structure of the SM is quite specific.
The gauge sector is invariant under a large global symmetry, 
\be
SU(3)_{Q_L}\times SU(3)_{D_R}\times SU(3)_{U_R}~,
\label{GF}
\ee 
corresponding to the family mixing of the
three independent fermion fields. 
This symmetry is broken only, and in 
in a well-defined way, by the Yukawa interaction.
The two Yukawa couplings $Y_U$ and $Y_D$
introduce breaking terms 
of the type 
\be
Y_U \sim 3_{Q_L} \times {\bar 3}_{U_R}~,  \qquad 
Y_D \sim 3_{Q_L} \times {\bar 3}_{D_R}~, 
\ee
which are highly hierarchical (with large 
entries only for the third family) and quasi aligned 
in the $SU(3)_{Q_L}$ sub-space (with the 
misalignment controlled by the off-diagonal 
entries of the CKM matrix).
This specific symmetry and symmetry-breaking structure 
is responsible for the successful SM predictions 
in the quark-flavour sector.

A natural and consistent possibility to export this 
pattern in the MSSM (as well as in other TeV-scale 
new-physics scenarios) is what  
goes under the name of Minimal Flavour Violation
(MFV) hypothesis~\cite{MFV}. According to this hypothesis, 
 $Y_U$ and $Y_D$ are the only breaking sources of 
the flavour symmetry also beyond the SM. 
As a result, the SM pattern for the suppression of FCNCs
is automatically fulfilled, and the constraints on the 
scale of new physics from rare processes do not exclude
the possibility of flavored degrees of freedom 
(the squarks) around or even slightly below 1~TeV. 

%
% Despite the MFV seems to be a natural solution to the flavour problem, 
% it should be stressed that we are still
% very far from having proved the validity of this hypothesis from data.
% A proof of the MFV hypothesis 
% can be achieved only with a positive evidence of physics beyond 
% the SM exhibiting the flavour pattern (link between $s\to d$, $b\to d$, and  
% $b\to s$ transitions) predicted by the MFV assumption~\cite{MFV}.
%

In most flavour observables the MFV hypothesis implies small ($\lsim 10\%$)
deviations from the SM (reason why this scenario is still consistent
with present data). However, as I will discuss in the rest 
of this talk, a notable exception is provided by helicity-suppressed 
observables in the large $\tan\beta$ regime of the MSSM. 
Here deviations from the SM at low energies can be large also under 
the MFV hypothesis, and improved data on low-energy observables 
turns out to be a key ingredient to identify this scenario.

\section{\boldmath MFV at large $\tan\beta$: general considerations}

The Higgs sector of the MSSM 
consists of two  $SU(2)_L$ scalar doublets, 
coupled separately to up- and down-type quarks
\bea
\cL^{\rm tree}_{\rm Y}  &=&  {\bar Q}_L {Y_U} U_R  H_U  
+ {\bar Q}_L Y_D D_R  H_D \no \\
&& + {\bar L}_L {Y_E} E_R  H_D  + V(H_U,H_D)+{\rm h.c.} \quad 
\label{eq:LYtree}
\eea
A key parameter of this sector is
the ratio 
of the two Higgs vevs:
$\tan\beta = \langle H_U \rangle /\langle H_D \rangle$.
Varying  $\tan\beta$
leads to modify the overall normalization 
of the two Yukawa couplings
(without changing their  misalignment in fla\-vour space).
It is therefore not surprising that this parameter plays a key role 
in flav\-our physics, especially under the MFV hypothesis.

On the theoretical side, the large $\tan\beta$ 
regime (or $\tan\beta=\cO(m_t/m_b) \approx 30-50$) has an 
intrinsic interest since it allows the unification of 
top and bottom Yu\-ka\-wa couplings, as predicted 
for instance in $SO(10)$ models of grand unification
(see e.g.~Ref.~\cite{GUT}). 
As recently stressed in Ref.~\cite{Hisano}, an independent 
motivation of large $\tan\beta$ is found in gauge-mediated
models.  Here the hierarchy $\langle H_U \rangle \gg  \langle H_D \rangle$
naturally follows from the tree-level conditions 
on the soft supersymmetry-breaking terms and the RGE 
evolution.
In models with gauge-mediated supersymmetry-breaking the 
flavour structure of the model is also naturally 
consistent with the MFV hypothesis. As a result, 
a MFV scenario with large $\tan\beta$ is not a
construction {\em ad hoc} to analyse interesting 
effects in flavour physics: these two conditions 
can both be realized in well-motivated
super\-sym\-metry-breaking scenarios.

Before discussing the implications of the large $\tan\beta$ 
regime in flavour-violating observables, it is 
worth discussing some general aspects 
of this scenario as well as its phenomenological 
motivations beside flavour phy\-sics.

\subsection{The effective Yukawa interaction at large $\tan\beta$ }
\label{sect:eff_Y}

\begin{figure}[t]
\begin{center}
\vspace{0.3 cm}
\includegraphics[scale=0.5]{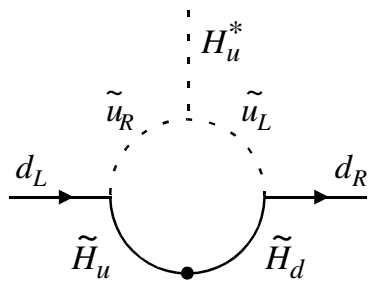} 
\end{center} 
\caption{\label{fig:HRS} Typical non-holomorphic coupling 
of the $H_U$ field to down-type quarks~\cite{HRS}.}
\end{figure}

The tree-level Lagrangian in Eq.~(\ref{eq:LYtree})
has an additional global symmetry with respect 
to the SM Yukawa interaction: it is invariant 
under an Abelian phase rotation of $H_D$ and $D_R$ 
fields with opposite charge (${\rm U}(1)_{\rm PQ}$ symmetry).
However, this symmetry cannot be an exact symmetry of the full MSSM 
Lagrangian: it has to be broken at least in the scalar potential
in order to avoid the presence of a massless pseudoscalar Higgs
boson. Beyond the tree-level quantum corrections transmit 
the ${\rm U}(1)_{\rm PQ}$ breaking to the Yukawa sector.
As a result, the effective Yukawa interaction obtained
by summing the leading quantum corrections may become 
substantially different from the one in Eq.~(\ref{eq:LYtree}). 

The most interesting phenomenon induced by 
the ${\rm U}(1)_{\rm PQ}$ breaking are effective 
non-holomorphic
couplings of the $H_U$ field to down-type quarks~\cite{HRS}
(generated by one-loop diagrams of the 
type in Fig.~\ref{fig:HRS} and similar diagrams with 
gluino exchange). Being generated only at the
quantum level, these effective couplings are small
 ($\epsilon \sim 1/(16\pi)^2$).
However, since the $H_U$ field has a large vev, 
the non-holomorphic terms induce large 
corrections to the vacuum structure of the theory 
in the large $\tan\beta$ 
regime ($\epsilon\tan\beta \sim 1$). 
Moreover, these one-loop amplitudes give rise to 
dimension-four effective operators, which 
are not suppressed in the limit of a heavy 
supersymmetry breaking scale. 

Two steps are necessary to re-sum the leading 
non-decoupling corrections to all orders~\cite{MFV,IR}:
\begin{itemize}
\item 
the structure of the effective Yukawa interaction 
($\cL^{\rm eff}_{\rm Y}$)
must be determined starting from 
the one-loop effective potential,  before determining
the vacuum structure of the theory
(i.e.~independently of the spontaneous breaking 
of the $SU(2)_L$ symmetry);
\item
the quark mass eigenstates must be defined by the 
diagonalization of the effective Yukawa interaction
at the minimum of the Higgs potential. 
\end{itemize}
By this way all the leading  $\cO(\epsilon~\tan\beta)$  terms 
are automatically included in the modified 
relations between Yukawa couplings
and physical observables. The most notable 
consequences are the redefinition of the diagonal
down-type Yukawa couplings in terms of quark masses~\cite{HRS}; 
a modification of the relation between CKM matrix
elements and off-diagonal Yukawa couplings~\cite{HRS};
a modified structure for the charged-Higgs 
couplings to quarks~\cite{BXsg1,BXsg2}.
Last but not least, a sizable FCNC coupling of down-type 
quarks to the heavy neutral Higgs fields ($H^0$, $A^0$) 
is generated~\cite{Babu,IR}.\footnote{The construction of 
$\cL^{\rm eff}_{\rm Y}$ and the identification 
of the leading $\cO(\epsilon~\tan\beta)$ terms 
is straightforward in the limit of the double 
hierarchy $M^2_{\rm sfermions} \gg M^2_H \gg M^2_W$, 
where we can neglect effective 
operators with dimension higher than six in $\cL^{\rm eff}_{\rm Y}$.
As discussed by Trine at this conference~\cite{Trine}, 
this method can be extended also to the less trivial 
case $M_H \sim M_W$: this extension has allowed to 
clarify the controversial claim about new types of 
large $\tan\beta$ effects in the $M_H \sim M_W$
region~\cite{Haish}. }

The latter phenomenon, which is particularly relevant for 
flavour physics, can easily be understood by noting that 
the diagram in  Fig.~\ref{fig:HRS}
generates an effective interaction 
of the type
\be
\delta \cL =  \epsilon_{\chi} {\bar Q}_L  Y_UY_U^\dagger Y_D D_R  (H_U)^c~.
\label{eq:L_nonholo}
\ee
This term is not aligned (in flavour space) and has  
a different composition of $h^0$, $H^0$, $A^0$, 
and vev components with respect to the 
leading term (${\bar Q}_L  Y_D D_R H_D$).
As a result, after the diagonalization of quark masses 
an effective FCNC coupling to the heavy  fields $H^0$
and $A^0$ is generated. The strength of this 
FCNC is controlled by the off-diagonal entries of
the CKM matrix, similarly to all the other FCNC
amplitudes in a MFV framework.
Being suppressed by the presence of $Y_D$, 
this effect is negligible in most FCNC amplitudes, 
but for the helicity-suppressed ones
(where also the SM contribution is suppressed 
by down-type masses). In the latter case the 
impact of Higgs-mediated FCNC amplitudes 
can be quite in the large: the FCNC coupling 
grows quadratically with $\tan\beta$
and is not suppressed in the limit of heavy 
squark masses.

\subsection{Phenomenological motivations in flavour-conserving processes}
\label{sect:mg_gm2}

\begin{figure}[t]
\begin{center}
\hspace{-0.3 cm}
\includegraphics[scale=0.36]{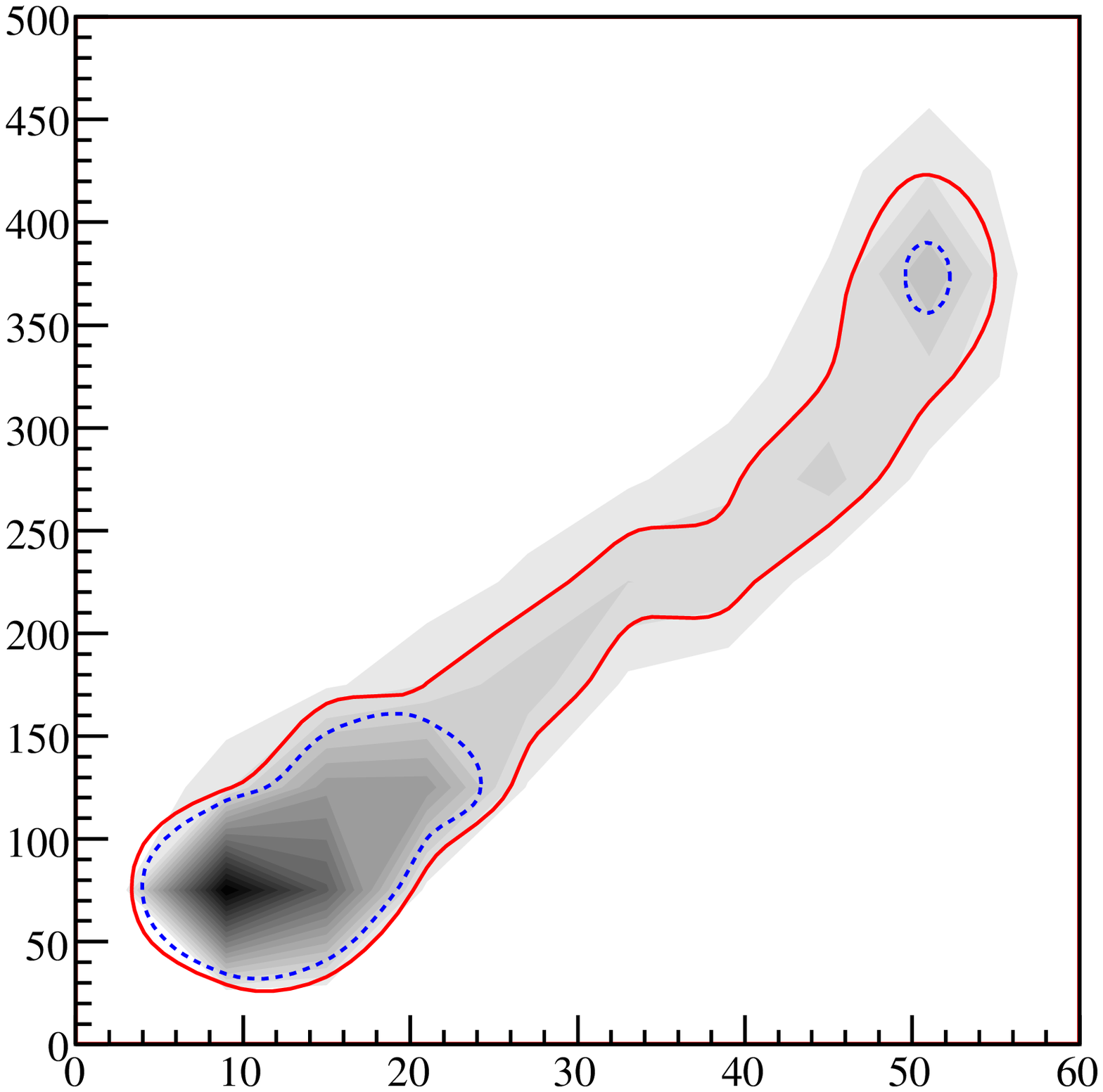} \\
\hspace{-0.3 cm}
\includegraphics[scale=0.49]{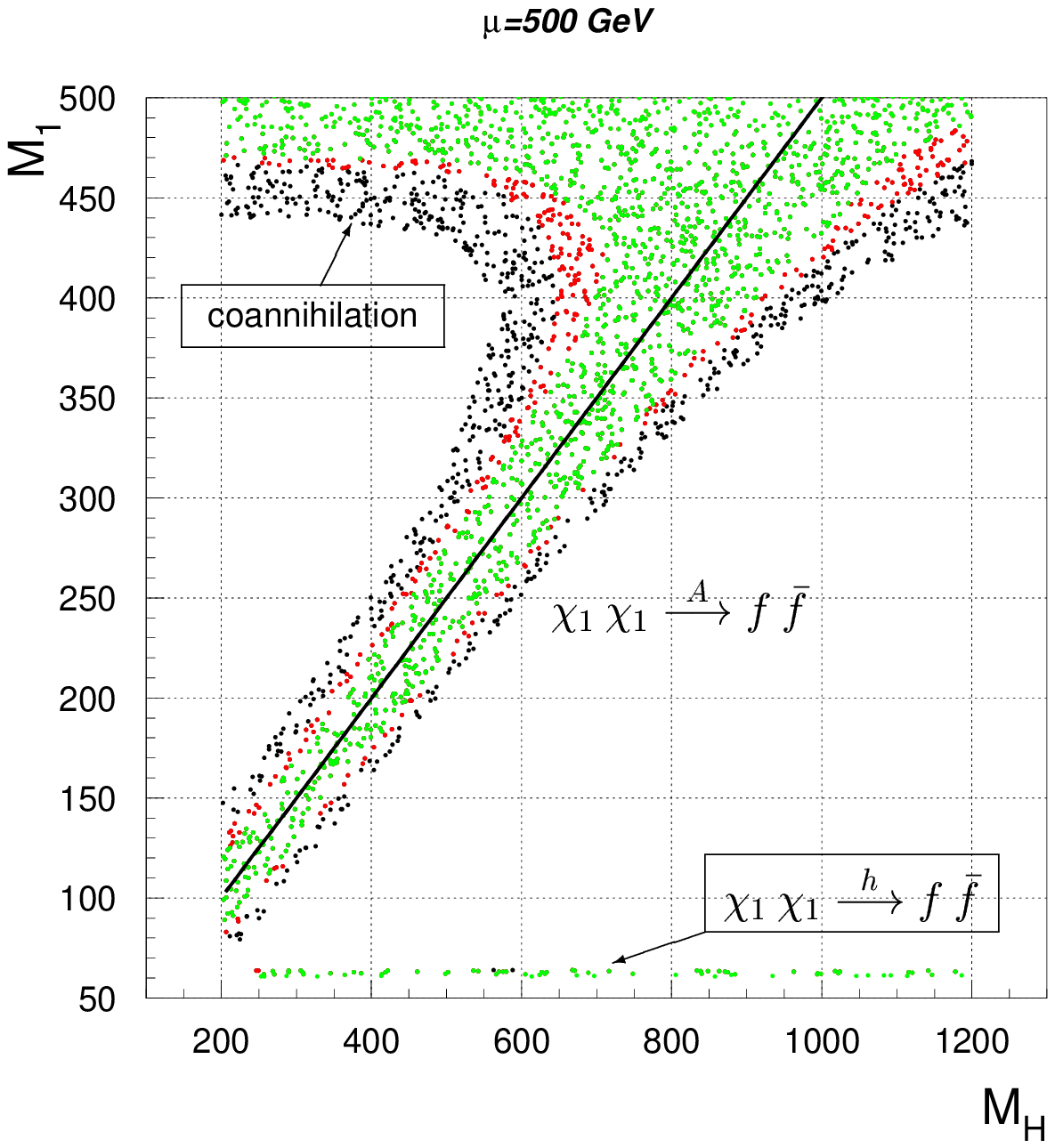}
\end{center}
\caption{\label{fig:Higgs} Up: Projection in the $M_H$--$\tan\beta$ 
plane of the global fit to the CMSSM~\cite{Oliver}.
Down: Allowed regions in the $M_1$--$M_H$ plane
satisfying the relic density constraint $\Omega h^2 < 0.119$
for $M_{\tilde q} = 2 M_{\tilde \ell} = |A_U| = 2\mu =1$~ TeV
and $\tan\beta=20$ (inner points), $30$ and $50$ (all points)~\protect\cite{Dark}. }
\end{figure}

The MFV scenario with large $\tan\beta$ 
is perfectly consistent with all the existing 
constraints from electroweak precision tests 
and flavour physics. An illustration of this
statement is provided by the plot in 
Fig.~\ref{fig:Higgs}, which is a
projection in the $M_H$--$\tan\beta$ 
plane of a recent global fit~\cite{Oliver} in 
the so-called constrained MSSM (CMSSM).
The parameter space of the CMSSM
is only a small subset of the general
MSSM with MFV. 
As shown in Fig.~\ref{fig:Higgs},
even within this restricted framework the 
corner with $\tan\beta =\cO(50)$ is 
well consistent with data and even 
slightly favored compared to other 
regions of the parameter space.

The shape of the plot in Fig.~\ref{fig:Higgs}
is modeled by two flavour-conserving 
observables, which can be inter\-pre\-ted as  
the first two hints of low-energy 
supersymmetry:  the anomalous 
magnetic of the muon and the neutralino 
relic abundance. As we will discuss below, 
in both cases large $\tan\beta$ values
are consistent and/or slightly favored 
by present data. 

\medskip

The possibility that the anomalous magnetic 
moment of the muon [$a_\mu = (g-2)_{\mu}/2$]
provides a first hint of physics beyond the SM 
has been widely discussed in the recent literature~\cite{gm2}.
As shown by Czarnecki~\cite{czar}, 
the consistency of the various  $e^+ e^-$  experiments 
in the determination of 
$(a_\mu)^{\rm SM}_{\rm had}$ has substantially increased 
our confidence in the SM prediction of this quantity.
As a result,  the discrepancy between the BNL measurement of 
$a_{\mu}$ and its SM prediction is now a solid 
$3$ sigma effect:
\be
 \Delta a_{\mu} =  a_{\mu}^{\rm exp} - a_{\mu}^{\rm SM}  
\approx (2.9 \pm 0.9) \times 10^{-9}~.
\label{eq:amu_exp}
\ee
The size of this discrepancy is about twice 
the electroweak SM contribution
($\Delta a_{\mu}^{\rm e.w.} \approx 1.5 \times 10^{-9}$).
Given the great success of the SM in the electroweak sector,
this fact is apparently very surprising. However, 
it can easily be explained noting that $a_\mu$ is an 
helicity suppressed observable, whose non-standard 
contribution can be enhanced compared to the 
SM one by increasing the value of $\tan\beta$.
Within the MSSM this enhancement can occur via 
gaugino-slepton loops, which generate a contribution 
to $a_\mu$ proportional to the muon Yukawa coupling
(and not to its mass)~~\cite{g_2_SUSY1}.
In the limit of almost degenerate 
soft-breaking terms, this can be written as
\be 
\Delta a^{\rm MSSM}_\mu \approx  \tan\beta  \times 
\Delta a_{\mu}^{\rm e.w.} \times 
\left( \frac{M_W}{\widetilde M_{\rm slept}} \right)^2~.
\ee
For values of $\tan\beta=\cO(10)$ the $M_W/{\widetilde M_{\rm slept}}$
suppression can easily be compensated for sleptons 
well above the $W$ mass, in prefect agreement with 
the constraints of electroweak precision tests.

\medskip 

Recent astrophysical observations consolidate the hypothesis
that the universe is full of dark matter localized in 
large clusters~\cite{wmap}. The cosmological density of 
this type of matter, which is likely to be composed by stable 
and  weakly-interactive massive particles, 
is determined with good accuracy 
\begin{equation}
0.079 \leq \Omega_{\rm CDM} h^2 \leq 0.119 \quad \rm{at}\: 2\sigma\:
\rm{C.L.}
\label{eq:omg}
\end{equation}
A perfect candidate for such form of matter is the lightest
neutralino of the MSSM (assuming $R$-parity conservation). 
In such case two key conditions need to be satisfied: 
i) the neutralino must be the lightest supersymmetric particle (LSP);
ii) it must have a sufficiently large annihilation cross-section
into SM matter.

The second condition, which is necessary given the large amount 
of neutralinos produced in the early universe compared the upper bound 
in Eq.~(\ref{eq:omg}), is not easily satisfied.
In most of the phenomenologically-allowed regions of the MSSM 
the lightest neutralino (usually a $B$-ino) has a very low annihilation 
cross section. At low values of $\tan\beta$ there are essentially two mechanisms
which can enhance this cross-section: i) light sfermions
(such that the $t$-channel sfermion exchange leads to a sufficiently 
large annihilation amplitude); 
ii) the co-annihilation with an almost degenerate NLSP.

The large $\tan\beta$ region has the virtue of allowing a third 
enhancement mechanism for the annihilation cross-section
of the relic neutralinos: the so-called $A$ funnel region. 
Here the dominant neutralino annihilation 
amplitude is the $s$-channel heavy-Higgs exchange. 
As illustrated in Fig.~\ref{fig:Higgs} (down), the size of the 
allowed region for this mechanism grows with  $\tan\beta$
and it can become very large for $\tan\beta =\cO(50)$.
This is the main reason for the local 
maximum around  $\tan\beta \approx 50$ in Fig.~\ref{fig:Higgs} (up).
As we will discuss in the following, 
Fulfilling the dark matter constraints via the 
$A$-funnel mechanism leads to well defined signatures 
in flavour physics. Indeed several of the parameters
which control the amount of relic abundance also play 
a key role in flavour observables~\cite{Dark,Lunghi}

\section{Large $\tan\beta$ effects in $B$ (and $K$) physics}
\label{sect:Bphys}

In the MFV scenario we are considering 
the overall normalization of $Y_D$  is largely enhanced 
compared to the SM case. However, its misalignment in 
flavour spa\-ce with respect to $Y_U$ is not modified. 
The latter property (following from the MFV ansatz)
implies that flavour-changing 
observables not suppressed by powers of down-type 
quark masses (i.e.~most of the experimentally accessible 
observables) are not sensitive to the value of $\tan\beta$. 
The interesting effects induced by $\tan\beta \gg 1$
show up only in the few observables sensitive to 
helicity-suppressed amplitudes.
These are confined to the $B$-meson system
(because of the large $b$-quark Yukawa coupling),
with the notable exception of $K\to\ell\nu$ decays.
We can divide the most interesting observables
in three classes: the charged-current processes $B(K) \to \ell \nu$,
the rare decays $B_{s,d} \to \ell^+\ell^-$, 
and the FCNC transition $B \to X_s \gamma$. 

\subsection{$B(K) \to \ell \nu$}

The charged-current processes $P \to \ell \nu$
are the simplest case. Here both SM and 
Higgs-mediated contributions (sensitive to $\tan\beta$) 
are dominated by a tree-level amplitude. 
The SM branching ratio can be written as
\bea
\label{eq:BR_B_taunu}
&& {\cal B}(P \to \ell \nu )^{\rm SM} =
\frac{G_{F}^{2}m_{P}m_{\ell}^{2}}{8\pi} \times \no\\
&& \qquad \left(1-\frac{m_{\ell}^{2}}
{m_{P}^{2}}\right)^{2}f_{P}^{2}|V_{uq}|^{2}\tau_{P} (1+\delta_{\rm e.m.})
\eea
where $V_{uq}=V_{ub}(V_{us})$ for $P=B(K)$
and $\delta_{\rm e.m.}$ denotes the electromagnetic 
corrections.

Within two-Higgs doublet models, the $H^{\pm}$--exchan\-ge amplitude 
induces an additional tree-level contribution to semileptonic decays
proportional to the Yu\-ka\-wa couplings of quarks and leptons~\cite{Hou}.
This can compete with the $W^{\pm}$ exchange only in 
$P \to \ell \nu$ decays, thanks to the 
helicity suppression of the SM amplitude.  
Taking into account the resummation of the leading $\tan\beta$
corrections to all orders, the $H^\pm$ contributions to 
the  $P \to \ell \nu $ amplitude within a MFV supersymmetric framework 
leads to the following ratio~\cite{IP,btnu}:
\bea
&& R_{P\ell\nu} = \frac{\BR( P\ell\nu)}{\BR^{\rm SM}( P\ell\nu)}  \no \\
&& \quad \stackrel{\rm SUSY}{=} \left[1-\left(\frac{m^{2}_P}{m^{2}_{H^\pm}}\right)
\frac{\tan^2\beta}{(1+\epsilon_0\tan\beta)}
\right]^2~,
\label{eq:Btn}
\eea
where $\epsilon_0$ denotes the effective coupling which 
parame\-trizes the non-holomorphic corrections to 
the down-type Yukawa interaction~\cite{Babu,IR}. 
For a natural choice of the MSSM parameters 
Eq.~(\ref{eq:Btn}) implies a suppression 
with respect to the SM in $B$ decays of
few$\times10\%$ (but an enhancement is also possible 
for very light $M_{H^\pm}$) and an effect 100
times smaller in $K$ decays (where the branching ratio
is necessarily smaller than $\BR^{\rm SM}$). 

\begin{figure}[t]
\centering
\includegraphics[scale=0.6]{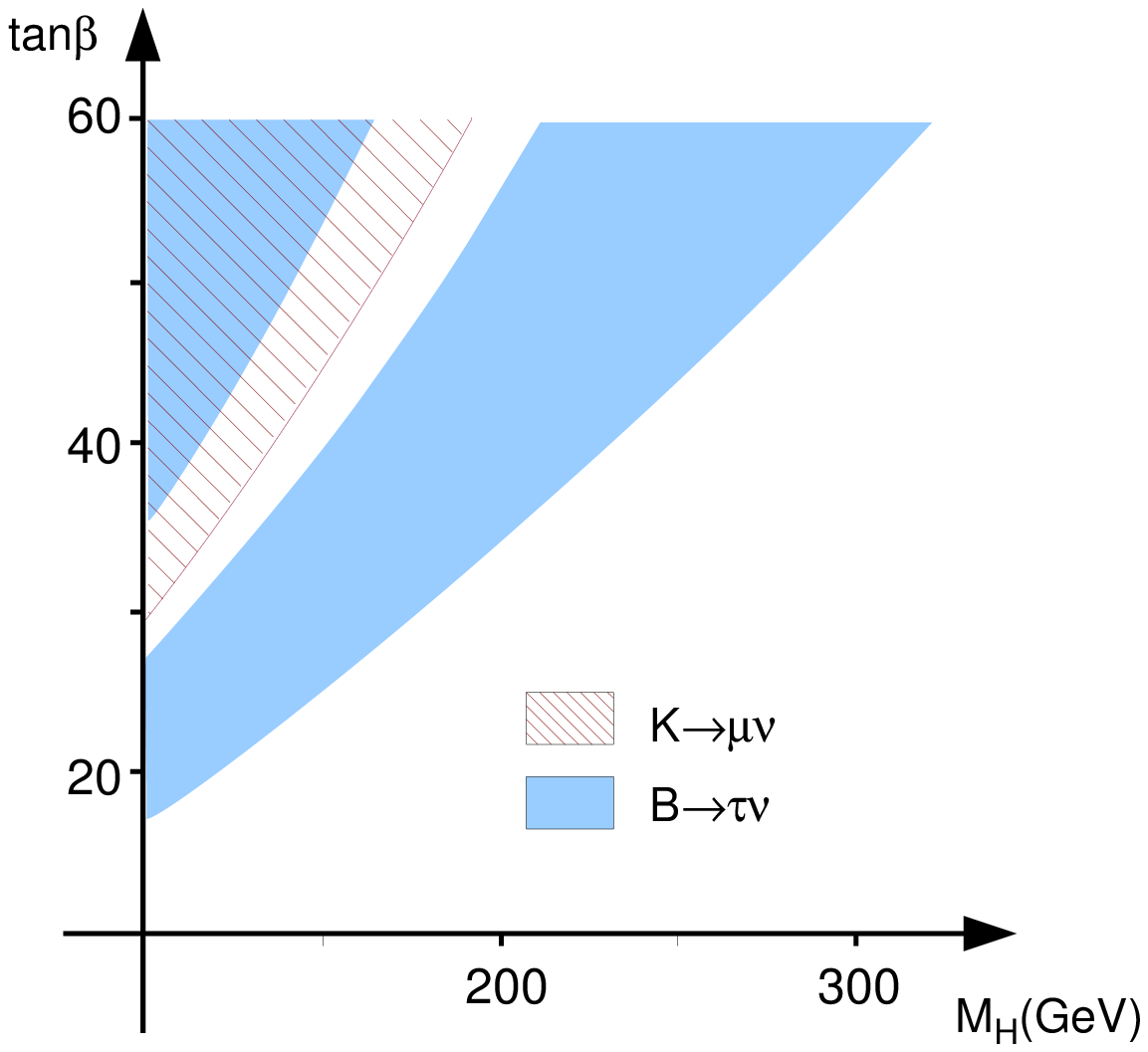} 
\vskip  0.5 cm
\caption{Present constraints in the $M_{H}$--$\tan\beta$ plane
from $\BR(\Btaun)$ and $\BR(K\to \mu\nu)$ \cite{Flavianet}.
\label{fig:Pln}}
\end{figure}

In the $B$ case only the $\tau$ modes has been observed. 
The average of the latest results by Babar  
and Belle yields \cite{Btaunu_exp} 
${\cal B}(\Btaun)^{\rm exp} = (1.41 \pm 0.43)\times 10^{-4}$
In the Kaon system both decay modes ($\ell=\mu,\nu$) 
are measured and the precision of
$\BR(K\to \mu\nu)$ is around $0.3\%$~\cite{KLOEkmn}.
Interestingly, the level of experimental precision 
in the combinations 
\beq
\frac{1}{m_B^2} [R_{B\to\tau\nu}-1] \qquad 
{\rm and} \qquad 
\frac{1}{m_K^2} [R_{K\to\tau\nu}]
\eeq
is comparable. 
In the limit of negligible theoretical errors 
we should therefore expect similar bounds 
in the  $M_{H}$--$\tan\beta$ plane from 
$B$ and $K$ decays. This limit is far from being 
realistic, due to the sizable errors on $f_P$ 
(determined from Lattice QCD) and $V_{uq}$ 
(which must be determined without 
using the information on $P \to \ell \nu$ decays).
But again the present level of precision is such that 
the $B$ and $K$ decays set competitive bounds in the 
$M_{H}$--$\tan\beta$ plane (see Fig.~\ref{fig:Pln}).

\subsection{$B\to \ell^+\ell^-$}

The important role of $\BR(B_{s,d} \to \ell^+ \ell^-)$
in the large $\tan\beta$ regime of the MSSM  has been widely 
discussed in the literature (see 
e.g.~Ref.~\cite{Lunghi,IP,Carena:2006ai,Ellis:2007fu}
for a recent discussion).
Similarly to $P\to \ell\nu$ decays, 
the leading non-SM contribution in 
$B\to \ell^+\ell^-$ decays is generated by a
single tree-level type amplitude:
the neutral Higgs exchange 
$B\to A,H \to \ell^+\ell^-$. 
Since the effective FCNC coupling of the neutral 
Higgs bosons appears only at the quantum level, in this 
case the amplitude has a strong dependence on 
other MSSM parameters in addition to  $M_{H}$ and $\tan\beta$.
In particular, a key role is played by $\mu$ and 
the up-type trilinear soft-breaking term ($A_U$),
which control the strength of the diagram 
in Fig.~\ref{fig:HRS}. The leading parametric 
dependence of the scalar FCNC amplitude from these 
parameters is given by 
\beq
\cA( B\to A,H \to \ell^+\ell^-) \propto \frac{m_b m_\ell}{M_A^2}
\frac{\mu A_U}{M^2_{\tilde q}} \tan^3\beta \times f_{\rm loop}
\eeq

\begin{figure}[t]
\centering
\vskip  0.3 cm
\hskip -1.5 cm
\includegraphics[scale=0.5]{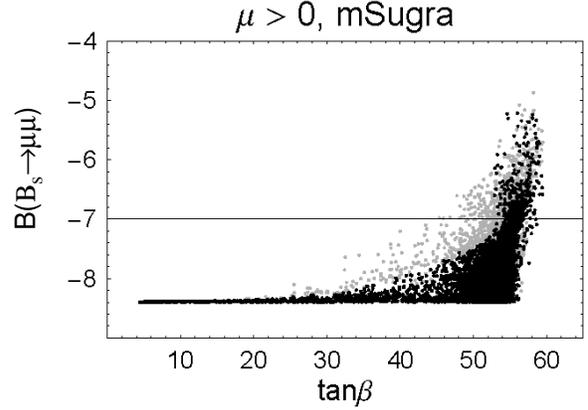} 
\vskip  0.3 cm
\caption{\label{fig:porod}  $\BR(B_s \to \mu^+\mu^-)$
as a function of $\tan\beta$ in the mSUGRA scenario~\cite{Lunghi}. }
\end{figure}

For $\tan\beta \sim 50$ and $M_A \sim 0.5$~TeV
the neutral-Higgs contribution to $\BR(B_{s,d} \to \ell^+ \ell^-)$
can easily lead to an $\cO(100)$ enhancement over the SM expectation.
This possibility is already excluded by experiments: the up\-per 
bound $\BR(B_s \to \mu^+\mu^-)< 5.8 \times 10^{-8}$~\cite{Bmm} 
is only about 15 times higher that the SM prediction~\cite{BmmSM}
\be
\BR^{\rm SM}(B_s \to \mu^+ \mu^-) = (3.4 \pm 0.5) \times 10^{-9}~. 
\label{eq:Bllexp}
\ee
This limit poses interesting constraints on the MSSM parameter space,
especially for light $M_H$ and large values 
of $\tan\beta$ (see e.g.~Fig.~\ref{fig:porod}). 
However, given the specific dependence on $A_U$ and $\mu$,
the present $\BR(B_s \to \mu^+\mu^-)$ bound does not exclude 
the large $\tan\beta$ effects in $(g-2)_\mu$ and 
$P\to\ell\nu$ already discussed. 
The only clear phenomenological conclusion which can be 
drawn for the present (improved) limit on $\BR(B_s \to \mu^+\mu^-)$
is the fact that the neutral-Higgs contribution 
to $\Delta M_{B_{s}}$ \cite{Buras}
is negligible.

\subsection{$B\to X_s \gamma$ }
The last flavour physics observable we need to
consider is $\BR(B\to X_s \gamma)$. As is well know, 
this FCNC transition is particularly sensitive to 
non-standard contributions, not only in the large
$\tan\beta$ regime of the MSSM. 
Contrary to pure leptonic decays discussed before, 
$B\to X_s \gamma$ does not receive effective 
tree-level contributions from the Higgs sector.
The one-loop charged-Higgs amplitude, which increases the
rate compared to the SM expectation, can be partially compensated 
by the chargino-squark amplitude, giving rise to delicate cancellations.  
As a result, the extraction of bound in the $M_H$--$\tan\beta$ plane
from $\BR(B\to X_s \gamma)$ (within the MSSM) 
is a non trivial task.  

Despite the complicated interplay of various non-standard 
contributions, $B\to X_s \gamma$ is particularly interesting 
given the good theoretical control of the SM prediction
and the small experimental error. 
According to the recent NNLO analysis of Ref.~\cite{bsgth},
the SM prediction is
\be
{\cal B}(B\to X_s \gamma)_{E_\gamma > 1.6~{\rm GeV}}^{\rm SM}
= (3.15 \pm 0.23) \times 10^{-4}~, 
\ee
to be compared with the experimental average~\cite{HFAG}:
\be
{\cal B}(B\to X_s \gamma)_{E_\gamma > 1.6~{\rm GeV})}^{\rm exp}
 = (3.55 \pm 0.24) \times 10^{-4}~. 
\ee
These results allow a small but non negligible
positive non-standard contribution to $\BR(B\to X_s \gamma)$
(as expected if the charged-Higgs amplitude would dominate over the 
chargino-squark one), which represents one of the most significant 
constraint  in the MSSM parameter space.

\begin{figure}[t]
\begin{center}
\includegraphics[scale=0.42]{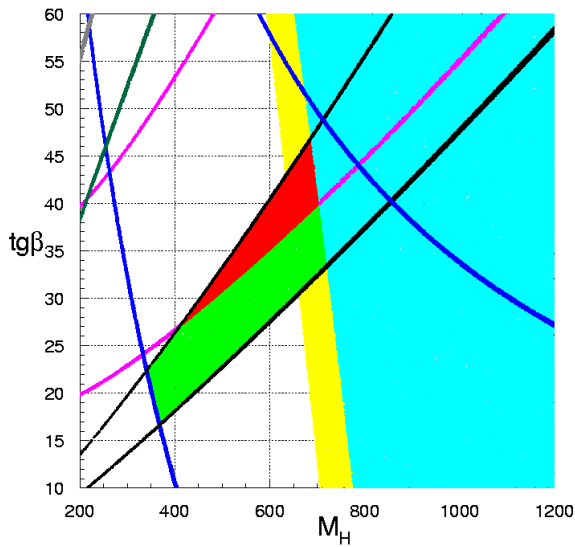} 
\end{center}
\caption{\label{fig:mu500}
Combined bounds from low-energy observables 
in the $\tan\beta$--$M_H$ plane assuming heavy squarks and 
dark-matter constraints in the $A$-funnel region~\protect\cite{IP}.
Main free parameters:
$M_{\tilde q} =1.5$~TeV, $A_U = -1$~TeV, $\mu=0.5$~TeV and 
$M_{\tilde{\ell}}=0.4$~TeV. Main low-energy constraints:
$1.01 < R_{ Bs\gamma} <1.24$ (region between the dark-gray (blue) 
lines falling at large $M_H$); 
$2 < 10^{9} (a_{\mu}^{\rm exp} - a_{\mu}^{\rm SM}) < 4$ 
(region between the two gray (purple) lines raising at large $M_H$);
$0.8 < R_{B\tau\nu} < 0.9$ (region between the two black lines 
raising at large $M_H$).
The light-gray (light-blue) area is excluded by the dark-matter 
conditions.
}
\end{figure}

\subsection{Discussion}
\label{sect:discuss}

The combined constraints on the three flavour observables 
discussed above leads to identify well-defined regions 
of the MSSM parameter space. For instance in the CMSSM
$\BR(B\to X_s\gamma)$  plays a key role, together with  $(g-2)_\mu$, 
in defining the preferred region showed in Fig.~\ref{fig:Higgs}.
As discussed in Ref.~\cite{NUHM}, in the NUHM scenario, 
where the universality condition between Higgs and 
sfermion soft breaking term is relaxed,
$\BR(B\to\tau\nu)$ and $\BR(B\to\mu^+\mu^-)$
are the most significant constraints in the 
light $M_H$ and large $\tan\beta$ region.

There also well-motivated scenarios where 
the pre\-sent constraints on this observables 
rule out most of the available parameter space.
As shown in Ref.~\cite{Albrecht:2007ii}, the present 
bound from $\BR(B\to\tau\nu)$ 
puts in serious difficulties the  SO(10) GUT model of 
Dermisek and Raby~\cite{Dermisek:2005ij}, which is 
a specific example of MFV scenario with large $\tan\beta$.

An illustration of the typical correlations 
of the low-energy constraints in the $M_H$--$\tan\beta$, 
in a generic scenario with heavy squarks and 
dark-matter condi\-tions satisfied in the 
$A$-funnel region, is shown in  Fig.~\ref{fig:mu500}.
The assumption of heavy squarks leads to a substantial 
simplification in the description of the large $\tan\beta$ 
effects, and in this regime we can 
draw the following general conclusions~\cite{Dark}:
1) The $B\to X_s \gamma$ constraint is always 
easily satisfied for $M_H \gsim 300$~GeV, or even 
lighter $M_H$  for large $\tan\beta$ values.  
This is because the present experimental range 
allows a significant (positive) non-standard contribution 
to the $B\to X_s \gamma$ rate, and choosing $A_U<0$
the positive charged-Higgs contribution is partially compensated
by the negative chargino-squarks amplitude. 
2) The present limit on  $B\to \mu^+ \mu^-$ is not particularly stringent.
3) A supersymmetric contribution to $a_\mu$
of $\cO(10^{-9})$ is perfectly compatible with the present constraints 
from $\BR(B \to X_s \gamma)$, especially for $A_U <0$.
Taking into account the correlation between 
neutralino and charged-Higgs masses occurring in the 
$A$-funnel region, this implies a 
suppression of  $\BR(\Btaun)$ with respect to 
its SM prediction of at least  $10\%$.
A more precise determination of $\BR(\Btaun)$ 
is therefore a key element to test this scenario.

\section{Lepton Flavour Violation and LF non-universality}
\label{sect:LFV}

LFV couplings naturally appear in the MSSM once we extend 
it to accommodate the non-vanishing neutrino masses 
and mixing angles by means of a supersymmetric seesaw mechanism~\cite{fbam}.
In particular, the renormalization-group-induced LFV entries 
appearing in the left-handed slepton mass matrices have the following 
form~\cite{fbam}:
\be
\delta_{LL}^{ij} ~=~ 
\frac{ \left( M^2_{\tilde \ell} \right)_{L_i L_j}}
{\sqrt{\left( M^2_{\tilde \ell} \right)_{L_i L_i} 
\left( M^2_{\tilde \ell} \right)_{L_j L_j}}} =
c_\nu (Y^\dagger_\nu Y_\nu)_{ij}~,
\label{eq:fbam}
\ee
where  $Y_\nu$ are the neutrino Yukawa couplings and 
$c_\nu$ is a numerical coefficient, depending 
on the SUSY spectrum, typically of  $\cO(0.1$--$1)$.
As is well known, the information from neutrino 
masses is not sufficient to determine in a model-independent 
way all the seesaw parameters relevant to LFV rates and,
in particular, the neutrino Yukawa couplings. 
To reduce the number of free parameters specific SUSY-GUT 
models and/or flavour symmetries need to be employed.
Two main roads are often considered in the literature:
the case where the charged-lepton LFV couplings are linked 
to the CKM matrix (the quark mixing matrix) and the case where 
they are connected to the PMNS matrix (the neutrino mixing 
matrix)~\cite{Arganda:2005ji}.
These two possibilities can be formulated in terms of
well-defined flavour-symmetry structures starting from the 
MFV hypothesis~\cite{MLFV,MFVGUT}.

Once non-vanishing LFV entries in the slepton mass matrices 
are generated, LFV rare decays are naturally induced by
one-loop diagrams with the exchange of gauginos and sleptons.
For large values of $\tan\beta$ the radiative decays
$\ell_{i}\rightarrow\ell_{j}\gamma$, mediated by 
dipole operators, are linearly enhanced, in close analogy to 
the  $\tan\beta$-enhancement of $\Delta a_\mu=(g_\mu-g^{\rm SM}_\mu)/2$.
A strong link between 
these two observable naturally emerges~\cite{hisano}.
We can indeed write
\bea
&& \frac{\BR(\ell_i\rightarrow \ell_j\gamma)}
{\BR(\ell_i\rightarrow \ell_j\nu_{\ell_i}\bar{\nu_{\ell_j}})} 
= \frac{48\pi^{3}\alpha}{G_{F}^{2}}
\left[\frac{\Delta a_{\mu}}{m_{\mu}^{2}}\right]^{2} \times \no \\
&& \qquad \times 
\left[\frac{f_{2c}\left( M^{2}_{2}/M^{2}_{\tilde \ell}, \mu^2/M^{2}_{\tilde \ell} \right)}{
g_{2c}\left( M^{2}_{2}/M^{2}_{\tilde \ell}, \mu^2/M^{2}_{\tilde \ell} \right)}
\right]^2
\,\left| \delta_{LL}^{ij} \right|^2~,
\label{eq:ratio_LFV}
\eea
where $f_{2c}$ and $g_{2c}$ are $\cO(1)$ loop functions. 
In the limit of a degenerate SUSY spectrum, this implies 
\beqa
&& \BR(\ell_i\rightarrow \ell_j \gamma) ~\approx~ 
\left[\frac{\Delta a_{\mu}}{ 20 \times 10^{-10}}\right]^{2} \times \no \\
&& \qquad \times 
 \left\{
\ba{ll}
1 \times 10^{-4} \, \left| \delta_{LL}^{12} \right|^2 \qquad  & [\mu\to e]~,  \\
2 \times 10^{-5} \, \left| \delta_{LL}^{23} \right|^2         & [\tau\to \mu]~.  
\ea
\right.
\label{eq:corr}
\eeqa

\begin{figure}[t]
\centering
\includegraphics[scale=0.40]{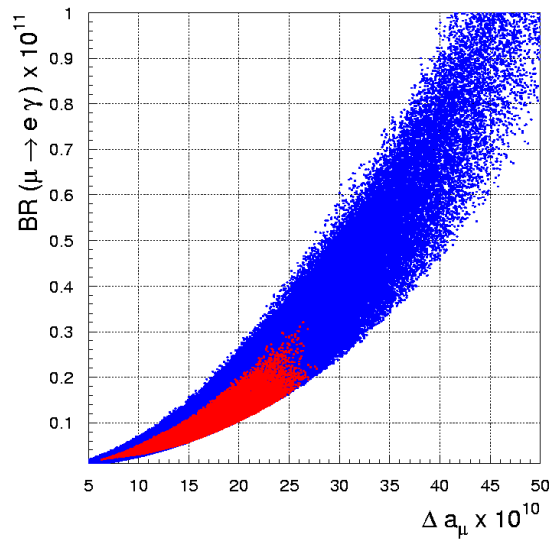} 
\caption{\label{fig2}
 $\cB(\mu\rightarrow e\gamma)$  
vs.~$\Delta a_\mu=(g_\mu-g^{\rm SM}_\mu)/2$, 
assuming $|\delta_{LL}^{12}|=10^{-4}$.
The scatter plot has been obtained employing the following ranges:
300~GeV~$\leq M_{\tilde{\ell}}\leq$~600~GeV, 200~GeV~$\leq M_{2}\leq$~1000~GeV,
500~GeV~$\leq \mu\leq$~1000~GeV, $10\leq \tan\beta \leq 50$, and 
setting $A_U=-1$~TeV, $M_{\tilde{q}}=1.5~$TeV.
Moreover, the GUT relations
$M_2\approx 2M_1$ and  $M_3\approx 6M_1$ are assumed.
The internal (red) area correspond to points within the 
$A$-funnel region~\protect\cite{IP} }
\end{figure}

The strong correlation between $\Delta a_{\mu}$ and the 
rate of the LFV transitions $\ell_i \rightarrow \ell_j\gamma$
holds well beyond the simplified assumptions used to 
derive these equations (see Fig.~\ref{fig2}).
The normalization $|\delta_{LL}^{12}|=10^{-4}$ used in 
Fig.~\ref{fig2} for  $\cB(\mu\rightarrow e\gamma)$
corresponds to the pessimistic MFV hypothesis in the 
lepton sector~\cite{MLFV}. As can be seen,
for such natural choice of $\delta_{LL}$ 
the $\mu\rightarrow e\gamma$ branching ratio is in the $10^{-12}$ 
range, i.e.~well within the reach of the MEG experiment~\cite{MEG}.

An independent and potentially large class of LFV 
contributions to rare decays in the 
large $\tan\beta$ regime of the MSSM 
comes from Higgs-mediated amplitudes. 
Similarly to the quark sector, 
non-holomorphic couplings can induce an effective FCNC 
Higgs coupling also in the 
lepton sector~\cite{Babu:2002et}.
Gauge- and Higgs-me\-dia\-ted amplitudes 
leads to very different 
correlations among LFV processes~\cite{Arganda:2005ji,Paradisi:2005tk,Paradisi:2006jp}
and that their combined study can reveal 
the underlying mechanism of LFV.

Finally, as recently pointed out in Ref.~\cite{Masiero:2005wr},
Higgs-mediated LFV effects at large $\tan\beta$ can also induce visible 
deviations of lepton-flavour universality 
in char\-ged-current processes. 
If the slepton sector contains 
sizable (non-minimal) sources of LFV, we could hope to observe 
deviations from the SM predictions in the 
$\BR(P\to \ell \nu)/\BR(P \to \ell^\prime \nu)$
ratios. The deviations can be $\cO(1\%)$ 
in $\BR(K\to e \nu)/\BR(K \to \mu \nu)$ \cite{Masiero:2005wr}, 
and  can reach $\cO(1)$ and $\cO(10^3)$ in  
$\BR(B\to \mu \nu)/\BR(B \to \tau \nu)$ and 
$\BR(B\to  e \nu)/\BR(B \to \tau \nu)$, respectively~\cite{IP}.

\section{Conclusions}
Within the Minimal Supersymmetric extension of the Standard Model,
the scenario with large $\tan\beta$ and Minimal Flavour Violation 
is well motivated and phenomenologically allowed. 
In this framework one could naturally accommodate 
the present (non-standard) central value of 
$(g-2)_\mu$, explain why the 
lightest Higgs boson has not been observed yet,
and why no signal of new physics has been
observed yet in $\BR(B\to X_s \gamma)$ 
and other flavour physics observables. 
Moreover, spectacular deviations from the SM 
in low-energy processes such as $B\to\mu^+\mu^-$
or $\mu\rightarrow e\gamma$ could be just around the corner. 

One of the interesting aspects of this scenario is the 
strong interplay between low-energy physics and
direct new-physics searches at high energy. As I tried to 
outline in this talk, 
improved measurements in the flavour sector,
particularly in the helicity suppressed decays 
$B(K) \to \ell \nu$, $B\to X_s \gamma$,
and $B\to\mu^+\mu^-$ represent a very useful 
tool to restrict the parameter space of the model, even
in absence of sizable deviations form the SM.

\section*{Acknowledgments} 
It is a pleasure to thank the organizers of 
SUSY 2007 for the invitation to this 
very interesting conference.  
This work is supported in part 
by the EU Contract No.~MRTN-CT-2006-035482 
{\em FLAVIAnet}.

\end{document}